\documentclass[aps,pra,twocolumn,showpacs]{revtex4-1}
\usepackage{amssymb}
\usepackage{graphicx}
\usepackage{dcolumn}
\usepackage{bm}
\usepackage{amsmath}
\usepackage{soul,color}
\usepackage{textcomp}
\usepackage{times}

\definecolor{cGreen}{RGB}{0,0,0}
\definecolor{cBlue}{RGB}{45,51,180}
\definecolor{cmagenta}{RGB}{205,0,100}

\usepackage[colorlinks,linkcolor=cBlue,citecolor=cBlue]{hyperref}

\begin{document}

\title{Learning Quantum Dissipation by Neural Ordinary Differential Equation}
\author{Li Chen$^{1,2}$}
\email{lchen@sxu.edu.cn}
\author{Yadong Wu$^{2,3}$}
\affiliation{
$^1${Institute of Theoretical Physics and State Key Laboratory of Quantum Optics and Quantum Optics Devices, Shanxi University, Taiyuan 030006, China}\\
$^2${Institute for Advanced Study, Tsinghua University, Beijing 100084, China}\\
$^3${State Key Laboratory of Surface Physics, Institute of Nanoelectronics and Quantum Computing, and Department of Physics, Fudan University, Shanghai 200433, China}
}

\begin{abstract}
Quantum dissipation arises from the unavoidable coupling between a quantum system and its surrounding environment, which is known as a major obstacle in the quantum processing of information. {Apart from its existence, how to trace the dissipation from observational data is a crucial topic that may stimulate manners to suppress the dissipation.} In this paper, we propose to learn the quantum dissipation from dynamical observations using the neural ordinary differential equation, and then demonstrate this method concretely on two open quantum-spin systems --- a large spin system and a spin-1/2 chain. We also investigate the learning efficiency of the dataset, which provides useful guidance for data acquisition in experiments. Our work promisingly facilitates effective modeling and decoherence suppression in open quantum systems.
\end{abstract}

\maketitle

\section{Introduction}
Quantum dissipation is closely related to such phenomena as decoherence, spectrum broadening and heating, all of which stand as serious obstacles in research areas ranging from quantum computation \cite{Ladd2010,Nielsen2010,Preskill2018} and simulation \cite{Georgescu2014,Bloch2012}, quantum information storage \cite{Lvovsky2009}, to quantum metrology \cite{Giovannetti2011,Pezze2018}, and sensing \cite{Degen2017}. The microscopic origin of dissipation is the breaking of isolation of the quantum system, i.e., the system inevitably interacts with its surrounding environment such that the information leakage occurs as the ambient degrees of freedom are traced out. Dissipation severely impairs the accuracy of modeling and manipulation of quantum systems.

Considerable efforts have been made to counteract the negative effects of dissipation. For example, quantum-nondemolition-mediated feedback has been used in suppressing the decoherence of a Schr{\"o}dinger cat state in a cavity \cite{Wiseman1993,Vitali1997}; spin-echo \cite{Hahn1950} and dynamical decoupling protocols \cite{Viola1999} have been widely applied to nitrogen-vacancy centers \cite{Lange2010,Du2009,Abobeih2018,Laraoui2013}, cold atoms \cite{Almog2010}, trapped ions \cite{Wang2017} and super-conducting circuits \cite{Guo2018,Pokharel2018}. These techniques often work with certain prior knowledge of the system-environment interactions. For example, in the nitrogen-vacancy centers, strong bias fields were introduced to circumvent the transverse coupling \cite{Du2009} or to effectively establish the bath correlations of the carbon nucleus \cite{Laraoui2013}. Furthermore, the coupling manner may also be recognized in advance, if it is magnetic or electric, linear or quadratical. {However, such prior information is generally unachievable, especially for many-body systems interacting with complex environments.}

{
In this paper, we propose a data-driven scheme to reconstruct the Markovian open quantum systems, where the dataset comes from the discrete observations of the relaxation dynamics under certain probes. Based on the dataset, we adopt the neural ordinary differential equation (NODE), a recently developed machine learning algorithm \cite{Chen2019}, to learn the Liouvillian of the open system by inversely solving the Lindblad (or Gorini-Kossakowski-Sudarshan-Lindblad) master equation \cite{Lindblad1976,Gorini1976,Gardiner2004}. A number of relevant works have been reported on such a topic. The operator expansion \cite{Franco2009} and the eigensystem realization algorithm (ERA) \cite{Zhang2014, Sone2017A} were firstly used in the time-trace of the Hamiltonians; the ERA was later extended to the Markovian dissipative systems  \cite{Zhang2015} and the estimation of the system size \cite{Sone2017B}. Several eigenstate- \cite{Qi2019, Bairey2019} or steady-state-based approaches \cite{Bairey2020} were further developed. A generic approach for local Hamiltonian tomography was reported using the initial and the final observations of quench dynamics \cite{Li2020}. Also for this topic, several traditional machine-learning architectures, such as the fully connected neural network \cite{Xin2019}, long short-term memory \cite{Che2021}, and convolutional neural network \cite{Ma2021} were recently involved. 

The novelty and contributions of our present work mainly lie in the following points. At first, our approach can be applied to either the Hamiltonian tomography (closed system) or the reconstruction of the Markovian dissipations (open system), with no need of prior information on the specific structures of the Hamiltonian or Liouvilian. Second, we adopt a relatively new machine-learning algorithm (namely the NODE \cite{Chen2019}) to deal with the gradient, with such issues encountered by the traditional machine-learning algorithms as the gradient explosion and vanishing \cite{Pascanu2012} being avoided. Furthermore, we also study the learning efficiency of datasets to facilitate the data acquisition in realistic experiments. We thus expect our work can play an active role in effective modeling and guiding new system-environment decoupling protocols for various open quantum systems.
}

\begin{figure}[t]
	\includegraphics[width=0.46\textwidth]{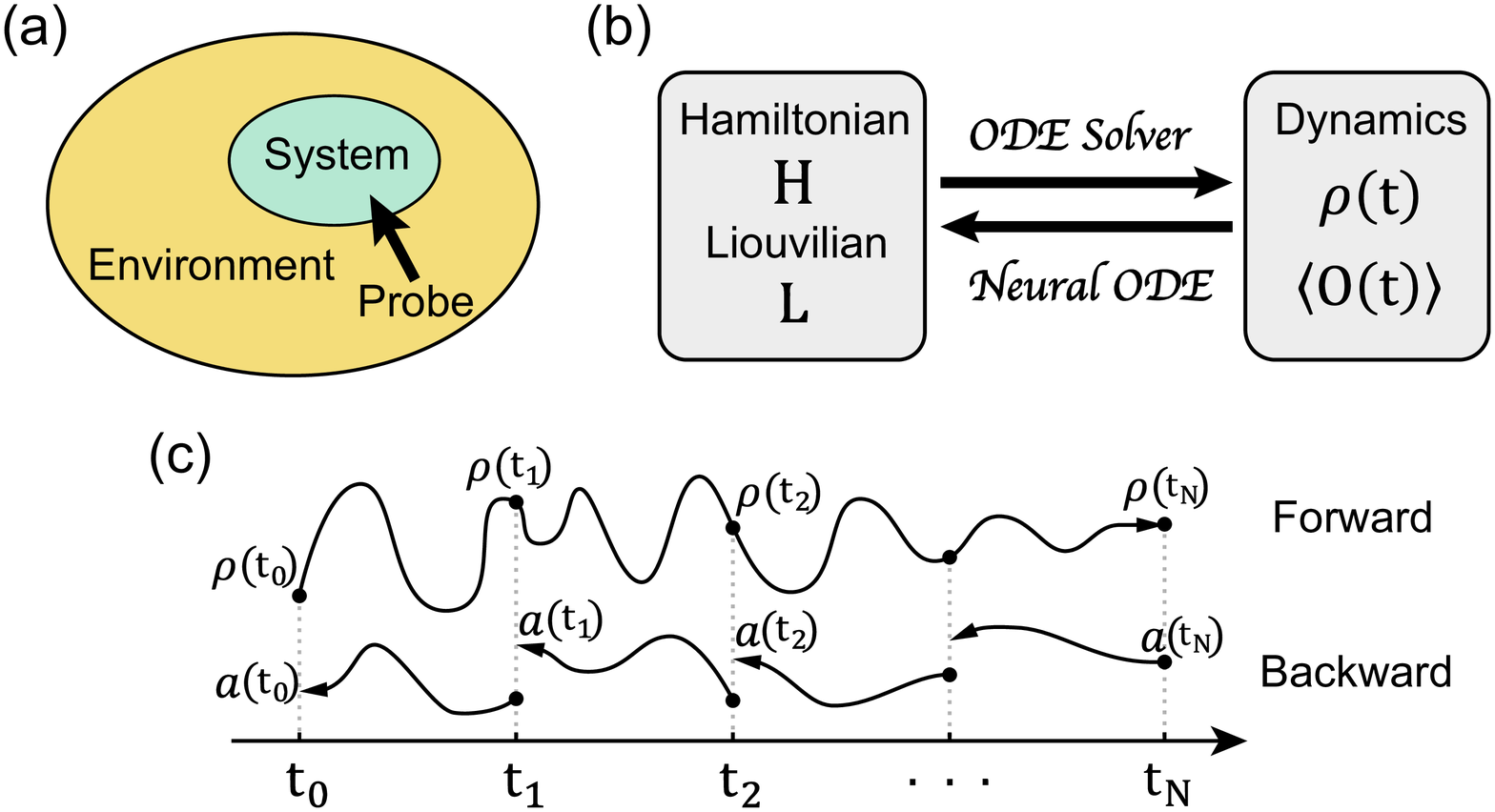}
	\caption{(a) Schematic of an open quantum system. (b) ODE solvers work to find solutions of the Lindblad equation when the Hamiltonian or the Liouvilian is priorly known, while the NODE works to reconstruct the Hamiltonian or Liouvilian as solutions ($\rho(t)$ or $\langle O (t)\rangle$) are given. (c) Mechanism of the NODE where $\rho(t)$ and $a(t)$ flow forward and backward, respectively. }
	\label{Fig1}
\end{figure}

\section{General Method}
We consider a quantum system being coupled to the environment, as is schematically displayed in Fig.~\ref{Fig1}(a). Under the Born-Markov approximation, the equation of motion of the system is governed by the Lindblad master equation \cite{Lindblad1976,Gorini1976,Gardiner2004}
\begin{equation}
\dot{\rho} = L[\rho],
\label{Liouvilian}
\end{equation}
where $\rho$ is the density matrix of the quantum system and $L$ is the Liouvilian super-operator in the form of (setting $\hbar=1$)
\begin{equation}
L[\rho] = -i[H,\rho] + \sum_k [J_k \rho J_k^\dagger - \frac{1}{2}\{J_k^\dagger J_k,\rho\}].
\label{GKSLE}
\end{equation}
Here, $H$ denotes the dissipation-independent Hamiltonian, $J_k$ is the dissipative operator (or jumping operator) in the $k$-th channel, and $\{J_k^\dagger J_k,\rho\} = J_k^\dagger J_k\rho + \rho J_k^\dagger J_k$ accounts for the normalization of $\rho$ as if no jump occurs.

Conventionally, with certain Liouvilian $L$ (namely $H$ and $J_k$), one can obtain the solution $\rho(t)$ by numerically solving the Lindblad equation (Eq.(\ref{GKSLE})) using ordinary differential equation (ODE) solvers such as the Euler or the Runge-Kutta \cite{Scherer2013}. In contrast, the goal of this paper is to determine the dissipation $J_k$ and even $H$ when the dynamical behaviors of density matrix $\rho(t)$ or certain observations $\langle O(t) \rangle = \text{Tr}[O\rho(t)]$ are given. In other words, we aim to deal with an inverse problem to reproduce $L$ from the sequential dataset $\mathcal{D} = \{\rho(t_n) \text{ or } \langle O(t_n) \rangle\}$, as illustrated in Fig.~\ref{Fig1}(b). Additionally, for the sake of experimental application, we consider that $\mathcal{D}$ is purely constructed by observations in the following discussion, i.e. $\mathcal{D} = \{\langle O(t_n) \rangle\}$, since compared to $\rho(t)$, acquiring $\{\langle O(t_n) \rangle\}$ is experimentally simpler and more feasible.

We adopt two different probes to generate the dataset $\mathcal{D}$. i) Time-dependent probe: for a fixed $\rho(t_0)$, one imposes a t-dependent control $H_\text{prob}(t)$, i.e., 
\begin{equation}
H = H_0 + H_\text{prob}(t),
\label{Hamiltonian}
\end{equation}
where $H_\text{prob} = \mathcal{P}(t)\cdot \mathbf{A}=\sum_\mu p_\mu(t) A_\mu$ with $p_\mu(t)$ being some smooth series and $A_\mu$ the according control operators. ii) Time-independent probe: with fixed $H$ and $J_k$, one diversifies $\langle O(t) \rangle$ by varying $\rho(t_0)$, which suits the experimental scenarios in which preparing different initial states is more convenient. For each evolution, we perform the measurement $\langle O(t_n) \rangle$ uniformly at discrete time within the time range $t_n\in [0,t_N)$, where $t_n = n t_N/N_\text{ts}$ with $N_\text{ts}$ being the time steps of measurement, which forms a data batch. The entire dataset is constructed by different batches, i.e., 
\begin{equation}
\mathcal{D} = \{\langle O(\mathcal{P}_s,t_n) \rangle| s\in [1,N_\text{bs}]; n\in [0,N_\text{ts}-1]\},
\label{dataset}
\end{equation}
where $N_\text{bs}$ denotes the batch-size. Therefore, the total number of data points in $\mathcal{D}$ is equal to $N = N_\text{bs} N_\text{ts}$. { Here, we mentioned that the discrete-time observations with equal intervals are considered for experimental convenience, which is however not a necessary condition for the learning algorithm that we will show below.}

To learn the Liouvilian from $\mathcal{D}$, we adopt a machine learning algorithm called the NODE \cite{Chen2019}. The NODE builds upon an ansatz $\dot{\rho}(\rho(t),t,\theta)$ with $\theta = \{\alpha,\gamma\}$, where $\alpha$ and $\gamma$ denote the parameters to be learnt in $H$ and $L_k$, respectively. The learning process is illustrated in Fig.~\ref{Fig1}(c), and it contains two parts. First, with $\dot{\rho}(\rho(t),t,\theta)$ and the initial state $\rho(t_0)$, we propagate the Lindblad equation (\ref{GKSLE}) forward by using certain ODE solvers, and then obtain a series of predictive solutions $\{\rho_\text{pred}(t_n)\}$; the loss function $\mathcal{L}$ is defined as a functional of $\langle O(t_n) \rangle_\text{pred}$ and $\langle O(t_n) \rangle_\text{real}\in\mathcal{D}$, which effectively measures the distance between $\mathcal{D}$ and the NODE predictions. The purpose of learning is to minimize $\mathcal{L}$  by adjusting $\theta$. To this end, the NODE introduces an adjoint field defined by $a(t) = \partial \mathcal{L}(t)/\partial \rho(t)$. Through backward propagating $a(t)$ from $t_N$ to $t_0$, we obtain $d\mathcal{L}/d \theta|_{t=t_0}$ which is then be used to update the parameters in the way of $\theta \rightarrow \theta - \lambda (d\mathcal{L}/d \theta|_{t=t_0})$ with $\lambda$ being the learning rate. More calculation details about the NODE algorithm can be found in Appendix \ref{SecI}.

\section{Examples}
We apply this method to two concrete examples. In the first example, we consider a one-body spin system with the spin quantum number being $S = 3/2$. Large spin systems are active in various research areas of quantum physics ranging from high-spin quantum dots \cite{Klochan2011,Doherty2013}, multi-component quantum gases \cite{Kawaguchi2012} to unconventional superconductors \cite{Wang2018}. In the second example, our system is a many-body spin-1/2 chain. Spin chains stand as fundamental models in condensed matter physics and quantum computation, which are closely related to quantum criticality \cite{Sachdev2011}, topological phase of matters \cite{Chiu2016,Wen2017}, etc.

\subsection{Spin-3/2 system} 
{The general spin-3/2 system refers to the four-level system with spin-vector operators $\{S_x,S_y,S_z\}$ being the generalized Pauli matrices. Since the spin-tensor operators may also be involved due to, for example, the quadratic Zeeman effect, we generally expand the Hamiltonian by the SU(4) generators, i.e.,
\begin{equation}
H_0 = -\sum_{\mu=1}^{15} \alpha_\mu \hat{g}_\mu,
\label{spin3/2-H}
\end{equation}
where $\hat{g}_\mu$ denote the 15 Hermitian generators satisfying traceless $\text{Tr}[\hat{g}_\mu]=0$ and orthogonal $\text{Tr}[\hat{g}_\mu \hat{g}_\nu]=2\delta_{\mu\nu}$ conditions, and $\alpha_\mu$ are the corresponding coefficients.} Furthermore, we assume that the system possesses weak dissipations in these Hermitian channels, namely
\begin{equation}
J_\mu = \sqrt{\gamma_\mu} \hat{g}_\mu,
\label{spin3/2-J}
\end{equation}
with $\gamma_\mu$ being the dissipative strength. The weakness is reflected in $\bar{\gamma} \ll \bar{\alpha}$, where $\bar{\alpha} = N_\alpha^{-1} \sum_\mu |\alpha_\mu|$ and $\bar{\gamma} = N_\gamma^{-1} \sum_\mu \gamma_\mu$ respectively denote the mean strength of $\alpha$ and $\gamma$, and $N_{\alpha,\gamma}$ are the number of parameters. Remark that our goal is to learn the parameters $\theta = \{\alpha,\gamma\}$ from the dataset $\mathcal{D}$.

The dataset $\mathcal{D}$ is generated by evolving the Lindblad equation Eq.~(\ref{Liouvilian}) within $t \in [0,20/\bar{\alpha})$ and by uniformly measuring the transverse magnetization, i.e., $\langle O \rangle = \langle S_x \rangle$ with $\{S_x,S_y,S_z\}$ the generalized Pauli operators. For general consideration, $\alpha$ and $\gamma$ are randomized within $\alpha\in [-1,1]\bar{\alpha}$ and $\gamma \in [0,2]\bar{\gamma}$, where $\bar{\gamma}=0.02\bar{\alpha}$. As mentioned before, we adopt two different probes, and for the t-dependent probe [probe i)], we fix $\rho(t_0) = \left|3/2,3/2\right\rangle \left\langle3/2,3/2\right|$ to be a magnetized state along $S_z$ and introduce a control magnetic field $H_\text{prob} = \sum_{\mu=x,y,z} p_\mu(t) S_\mu$ with $p_\mu(t) = 2\bar{\alpha}\sum_{m=1}^{10} \sin(2\pi \omega_m t)$ and $\omega_m$ being randomized within $\omega_m \in [-1,1]\bar{\alpha}$; for the t-independent probe [probe ii)], we take $\rho(t_0)$ to be random pure states that satisfy $\text{Tr}[\rho^2(t_0)]=1$. The total data numbers of $\mathcal{D}$ is $N = 10^3$ with $N_\text{bs}=10$ and $N_\text{ts}=10^2$.

\begin{figure}[t]
\includegraphics[width=0.48\textwidth]{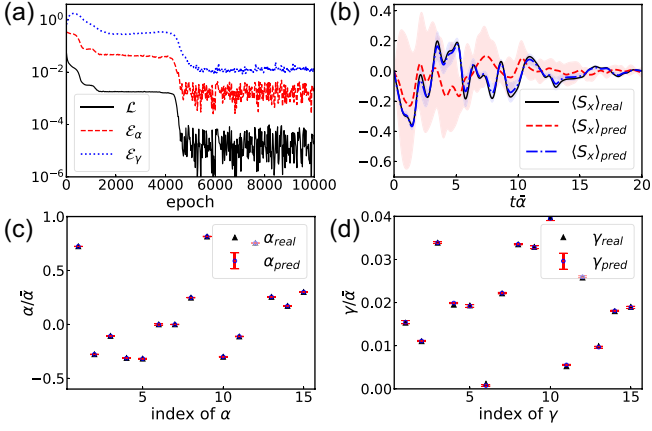}
\caption{Learning results of the spin-3/2 model with the t-dependent probe. (a) Dependence of the loss function $\mathcal{L}$, the relative error of $\alpha$, and the relative error of $\gamma$ on training epochs. (b) Spin dynamics where solid, dashed and dot-dashed lines correspond to $\langle S_x \rangle_\text{real}$ in the dataset $\mathcal{D}$, $\langle S_x \rangle_\text{pred}$ before training and $\langle S_x \rangle_\text{pred}$ after $10^3$ epochs of training. (c) and (d) Real parameters (solid triangles) and predictive parameters (circles with error bars) after training, respectively. Color shadings in (b) and error bars in (c) and (d) indicate the predictive fluctuations due to $N_\text{ini}=10$ different initializations. }
\label{Fig2}
\end{figure}

In the learning process, we minimize the mean square error (MSE) loss function
\begin{equation}
\mathcal{L} = \frac{1}{N}\sum_{s,n}\left[\langle S_x(\mathcal{P}_s,t_n) \rangle_\text{pred} - \langle S_x(\mathcal{P}_s,t_n) \rangle_\text{real}\right]^2,
\label{loss}
\end{equation}
and monitor the averaged relative error with respect to the parameters, i.e.,
\begin{equation}
\mathcal{E}_{\theta=\{\alpha,\gamma\}} =  \overline{\frac{1}{\bar{\theta}}\sqrt{\frac{1}{N_\theta}\sum (\theta_\text{pred} - \theta_\text{real})^2}},
\label{RE}
\end{equation}
where the additional overscore on the right-hand side means to take a further average on $N_\text{ini}$ different initializations of $\theta$. Here, we take $N_\text{ini} = 10$.

In Fig.~\ref{Fig2}, we present the learning results of $\mathcal{D}$ using the t-dependent probe, while leave those using the t-independent probe in Appendix \ref{SecII}. Specifically, Fig.~\ref{Fig2}(a) shows the variation of $\mathcal{L}$, $\mathcal{E}_\alpha$ and $\mathcal{E}_\gamma$ on training epochs, from which one can read that the algorithm converges at approximately $5\times 10^3$ epoch. After convergence, we compare the predictive parameters $\alpha$ and $\gamma$ (labeled by circles with error bars) with their realistic values (solid triangles) in subfigures (c) and (d), respectively, where the error bars indicate the standard deviation because of different initializations. Clearly, the algorithm successfully reproduces the parameters with relative errors $\mathcal{E}_\alpha \lesssim 5\text{\textperthousand}$ and $\mathcal{E}_\gamma \lesssim 2\%$. Fig.~\ref{Fig2}(b) presents the spin dynamics $\langle S_x (t)\rangle$ of a typical batch, where the solid, dashed and dot-dashed lines correspond to the realistic $\langle S_x (t)\rangle_\text{real}$, the predictive curve before training, and the predictive one at the $10^3$ training epoch respectively, with shadings indicating the predictive fluctuations due to initializations. As the training progresses, $\langle S_x \rangle_\text{pred}$ gradually approaches $\langle S_x \rangle_\text{real}$ accompanied by the diminishment of predictive fluctuations.

\subsection{Spin-1/2 Chain}
\begin{figure}[t]
\includegraphics[width=0.48\textwidth]{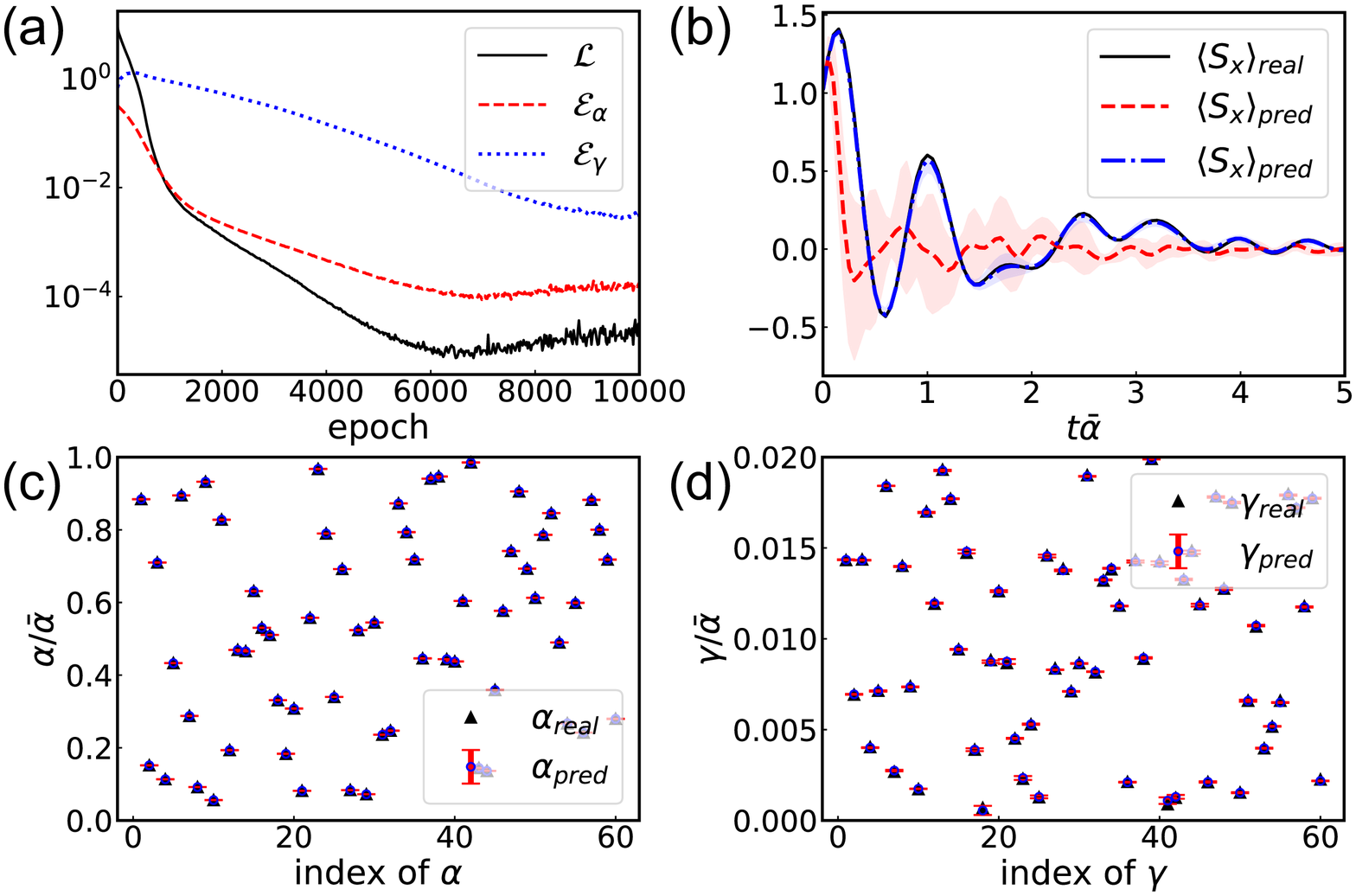}
\caption{Learning results of the spin-1/2 chain with the t-independent probe. (a) Dependence of $\mathcal{L}$, $\alpha$, and $\gamma$ on training epochs. (b) Total spin dynamics of $\langle S_x \rangle_\text{real} \in \mathcal{D}$, $\langle S_x \rangle_\text{pred}$ before training and $\langle S_x \rangle_\text{pred}$ after $10^3$ epochs of training. (c) and (d) Real parameters (solid triangles) and predictive parameters (circles with error bars) after training.}
\label{Fig4}
\end{figure}
The second example is a spin-1/2 chain with nearest-neighbor interactions, whose Hamiltonian is written as
\begin{equation}
H_0 = -\sum_{i,\mu} \alpha_\mu^i \sigma_\mu^i - \sum_{i,\mu,\nu} \alpha_{\mu\nu}^i \sigma_\mu^i \sigma_\nu^{i+1}.
\label{spin1/2-H}
\end{equation}
The first term accounts for the local terms with $\sigma_{\mu=x,y,z}^i$ the spin-1/2 Pauli operators with $\alpha_{\mu}^i$ the corresponding strength. The second term characterizes the two-body interactions with strength $\alpha_{\mu\nu}^i$. The one-body and two-body dissipations are considered in the according channels
\begin{equation}
\begin{aligned}
J_\mu^i &= \sqrt{\gamma_\mu^i} \sigma_\mu^i, \;\;\;\;\; J_{\mu\nu}^i &= \sqrt{\gamma_{\mu\nu}^i} \sigma_\mu^i \sigma_\nu^{i+1},
\label{spin1/2-J}
\end{aligned}
\end{equation}
with $\gamma_\mu$ and $\gamma_{\mu\nu}$ the corresponding strength. As a proof-of-principle demonstration, we practically set the chain length to 5 under a periodical boundary condition, which keeps the numerical complexity within the computational power of a PC with two GPUs \cite{Complexity}. In such a case, there is a total of 120 parameters that need to be learned, 60 each for $\alpha$ and $\gamma$. Again, we generate two datasets using the t-dependent and the t-independent probes. For the former, we set $\rho(t_0) = \bigotimes_i \left|\uparrow\right\rangle_i \left\langle\uparrow\right|_i$ and $H_\text{prob} = \sum_\mu p_\mu S_\mu$ with $S_{\mu=x,y,z} = \sum_i \sigma_\mu^i$ the total spin operators and $p_\mu = 2\bar{\alpha}\sum_{m=1}^{10} \sin(2\pi \omega_m t)$; for the latter, $\rho(t_0)$ are chosen to be random product states. The measured observables are the total spins, i.e., $\langle O \rangle = \langle S_\mu \rangle$. More detailed settings are listed below: $t\in [0,5/\bar{\alpha}]$, $N = 10^3$ with $N_\text{bs} = 20$ and $N_\text{ts} = 50$, randomized $\alpha\in[0,1]\bar{\alpha}$, $\gamma\in[0,0.02]\bar{\alpha}$, $\omega_m\in[-2,2]\bar{\alpha}$, and $N_\text{ini} = 10$.

In Fig.~\ref{Fig4}, we display the learning task with the t-independent probe while leaving the other task (using t-dependent probe) in Appendix \ref{SecII}. All instructions of the Fig.~\ref{Fig4} are similar to those used in Fig.~\ref{Fig2} except that $\langle S_x \rangle$ now denotes the total magnetization. The accurate predictions with $\mathcal{E}_\alpha \lesssim 1\text{\textperthousand}$ and $\mathcal{E}_\gamma \lesssim 1\%$ shown in Figs.~\ref{Fig4}(c) and (d) clearly demonstrate the feasibility of the algorithm on many-body spin systems.

{ Here, we would like to make several additional comments. At first, the system shown above carries no symmetry, which in principle allows us to learn all the parameters by looking at a global (or local) observable $\langle O \rangle$. However, if the system intrinsically carries certain symmetry, then all the accessible operators associated with $O$ may be limited within certain subspaces, which forms the accessible set \cite{Zhang2014}. In this case, only the parameters related to the accessible set can be determined by the current time-trace approach. Second, there is a case in which the NODE fails to make unique predictions even for systems without intrinsic symmetry. The NODE can access to three parts of information --- the initial state $\rho_0$, the probe field $H_\text{prob}$, and the measured data $\langle O \rangle \in \mathcal{D}$, and hence it cannot make unique predictions if all the three parts share a common "symmetry".} For example in the learning task using t-dependent probe (see Appendix \ref{SecII}), $\rho(t_0)$, $H_\text{prob}$ and $\langle S_x \rangle$ are all translational invariant (namely independent on local spins), which leads to unfavorable learning results because of the "symmetry"-induced ambiguity. This problem, however, does not occur for the task with the t-independent probe illustrated in Fig.~\ref{Fig4}, since the employed initial states $\rho(t_0)$ explicitly break the translational "symmetry".

\section{Learning Efficiency} 
In the above, we illustrated the capability of the algorithm in learning open quantum systems.
Now, we turn to the question of learning efficiency --- how to collect the data points can make the learning more efficient? We emphasize the importance of this question on data acquisition in realistic experiments, especially when collecting data points is expensive or time-consuming.

\begin{figure}[t]
\includegraphics[width=0.48\textwidth]{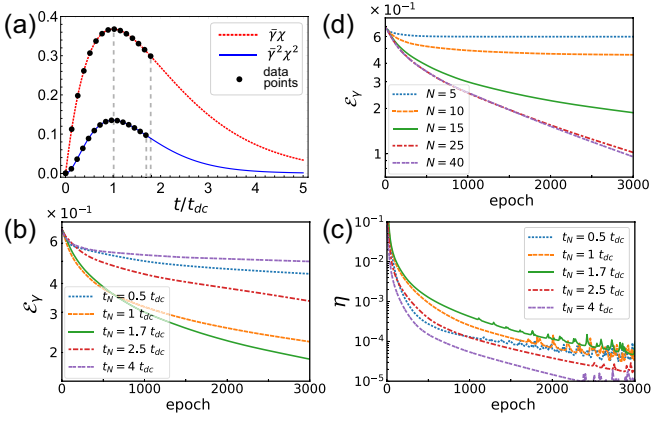}
\caption{(a) $\chi$ and $\chi^2$ as a function of $t$, where dots visually indicate the optimal strategy of data acquisition with $t_N=t_N^\text{opt}$. (b)-(d) learning results of the spin-3/2 example using the t-dependent probe and the MSE loss function. (b) and (c) respectively show the variation of $\mathcal{E}_\gamma$ and $\eta$ on $t_N$ with $N=N_\gamma=15$ being fixed. (d) shows the variation of $\mathcal{E}_\gamma$ on $N$ with $t_N=t_N^\text{opt}=1.7 t_\text{dc}$ being fixed. All the numerical results have been averaged on $N_\text{ini}=50$ random initializations.}
\label{Fig5}
\end{figure}

To make this question simple and clear, let us focus on the situation that $\gamma$ are the only unknown parameters that we would like to learn. Moreover, since different batches are independent of each other in $\mathcal{D}$, we consider $\mathcal{D}$ only contains one batch such that the total number of data points is $N=N_\text{ts}$.
As mentioned before, the NODE adjusts parameters $\gamma_\mu$ according to $d\mathcal{L}/d\gamma_\mu$, which motivates us to define
\begin{equation}
\eta = \frac{1}{NN_\gamma} \sum_{n=1}^N\sum_{\mu=1}^{N_\gamma} \left|\frac{d\mathcal{L}_n}{d\gamma_\mu}\right|,
\label{EP0}
\end{equation}
where $\mathcal{L}_n$ means the local loss function with respect to a single data point $\langle O(t_n) \rangle$. The physical meaning of $\eta$ is quite clear, it characterizes the averaged sensitivity of $\mathcal{L}$ on $\gamma$ such that a large $\eta$ would speed up the learning process. To further simplify the Eq.~(\ref{EP0}), we replace the average of individual derivatives by the derivative to the mean value of $\gamma_\mu$, i.e.,
\begin{equation}
\eta = \frac{1}{N} \sum_{n=1}^N \left|\frac{d\mathcal{L}_n}{d\bar{\gamma}}\right|.
\label{EP}
\end{equation}
For weak dissipation $\bar{\gamma} \ll \bar{\alpha}$, $\eta$ is closely related to
\begin{equation}
\begin{aligned}
\chi(t) &= \left|\frac{d\langle O(t) \rangle}{d\bar{\gamma}}\right| = \frac{\Delta_1 t e^{-\Delta_1 t}}{\bar{\gamma}} |f(t)|,
\end{aligned}
\label{Sus}
\end{equation}
which is composed of two parts: $|f(t)|$ is a fast-oscillating term characterized by the energy scale of $\bar{\alpha}$, while $t e^{-\Delta_1 t}$ is a slowly varying envelop depicting the dissipation-induced damping of $\langle O(t) \rangle$, where $\Delta_1 \propto \bar{\gamma}$ is the real part of the Liouvilian gap. Obviously, the envelop of $\chi(t)$ is a non-monotonic function being maximized at $t = t_\text{dc}$ (see $\chi(t)$ and the $\chi^2(t)$ in Fig.~\ref{Fig5}(a)), with $t_\text{dc} = 1/\Delta_1$ being the decoherent time.

The most commonly used two loss functions, mean absolute error (MAE) and MSE, are linearly and quadratically proportional to $\chi(t)$ as $\bar{\gamma}$ approaches its realistic value $\bar{\gamma}_\text{real}$, i.e.,
\begin{equation}
\begin{aligned}
\eta_\text{MAE} &= \frac{1}{N} \sum_{n} \chi(t_n), \\
\eta_\text{MSE} &\propto \frac{1}{N} \sum_n \chi^2(t_n),
\end{aligned}
\label{EPs}
\end{equation}
where MAE is defined by $\mathcal{L}_n = |\langle O(t_n) \rangle_\text{pred} - \langle O(t_n) \rangle_\text{real}|$, while the definition of MSE is presented in Eq.~(\ref{loss}).
For uniform measurement $t_n = nt_N/N_\text{ts}$, the summations in Eq.~(\ref{EPs}) can be analytically calculated in the large $N$ limit, i.e., ($t_N$ in the unit of $t_\text{dc}$)
\begin{equation}
\begin{aligned}
\lim_{N\rightarrow \infty}\eta_\text{MAE} &= \frac{1-e^{-t_N}-t_N e^{-t_N}}{t_N}, \\
\lim_{N\rightarrow \infty}\eta_\text{MSE} &\propto \frac{1-e^{-2 t_N}-2 t_N(t_N+1)e^{-2 t_N}}{4 t_N},
\end{aligned}
\label{EPs2}
\end{equation}
where both are non-monotonic with maximal values lying at $t_N^\text{opt} \approx 1.8t_\text{dc}$ and $t_N^\text{opt}\approx 1.7t_\text{dc}$, respectively. This indicates that the optimal strategy of uniform data acquisition is to take  $t_N=t_N^\text{opt}$, as is schematically illustrated by dots in Fig.~\ref{Fig5}(a).
We benchmark the above analysis on the spin-3/2 learning task using the t-dependent probe and the MSE loss. This task is with $t_\text{dc}\approx 9/\bar{\alpha}$. Practically, we set $N=15$ and plot the variations of $\mathcal{E}_\gamma$ and $\eta$ in Figs.~\ref{Fig5}(b) and (c) respectively, where $\eta$ is numerically calculated using Eq.~(\ref{EP0}). Clearly, $t_N=t_N^\text{opt}=1.7 t_\text{dc}$ exhibits the largest $\eta$ and the fastest decay rate of $\mathcal{E}_\gamma$ throughout the learning process, which is in good agreement with our previous discussion.

Finally, we briefly discuss the effect of the total data number $N$. Therefore, we fix $t_N=t_N^\text{opt}$ and show the dependence of $\mathcal{E}_\gamma$ on $N$ in Fig.~\ref{Fig5}(d). An apparent feature is that the cases with $N < N_\gamma = 15$ exhibit poor learning results, which is understandable since we need at least $N_\gamma$ data points to uniquely determine all $\gamma$, resembling that one needs at least $N_\gamma$ equations to determine $N_\gamma$ variables. Furthermore, an increasing $N$ can accelerate the learning process, however, this trend will not continue endlessly (comparing the curves $N=15$, $N=25$ and $N=40$ in Fig.~\ref{Fig5}(d)). For a fairly large $N$, neighboring data points exhibit tiny distinguishment such that adding more data points would be of little help for the learning efficiency.

\section{Summary and Outlook}
We proposed a scheme to learn the quantum dissipation of open systems based on the NODE, a machine learning algorithm being able to reproduce the Liouvilian from dynamical observations. The learning process can be accelerated by optimizing the strategy of data collection. There are many follow-up questions. {The Lindblad master equation relies on the Born-Markov approximation, which limits the current method to weakly dissipative systems with short correlation time. The generalization from Markov to non-Markov is not so straightforward, since the memory effect leads to an integro-differential master equation \cite{Gardiner2000,Vega2017}. More advanced techniques are expected to deal with the gradients in relation to the integral kernel. Whether an effective Markovian description can be found for non-Markovian dynamics is still left open. 
Furthermore, the full-Liouvilian calculation shown above cannot be easily extended to large-scale many-body systems due to the exponential growth of the Hilbert space. One possible solution is to combine the NODE with quantum-trajectory approaches \cite{Gardiner2000, Daley2014} such as the truncated Wigner method \cite{Schachenmayer2015,Huber2022} and the tDMRG-quantum-trajectory method \cite{Daley2009}. In this regard, several recently developed machine-learning-based solvers \cite{Carleo2017, Nagy2019, Hartmann2019, Vicentini2019, Mazza2021, Liu2022, Luo2022} may provide valuable insights.}
Additionally, with a dissipative model, can machine learning algorithms help design the corresponding protocols for system-environment decoupling? We expect this work, as well as these questions, to stimulate more interdisciplinary studies in the fields of machine learning and open quantum systems.

\begin{acknowledgments}
L.C. would like to thank Hui Zhai, Ce Wang, Juan Yao, Lei Pan, Yang Shen, and Sen Yang for the fruitful discussion. L.C. acknowledges support from the National Natural Science Foundation of China (Grant Nos. 12174236 and 12147215) and the postdoctoral fellowship offered by Hui Zhai. Part of this work was done during the fellowship at Tsinghua University, Beijing.
\end{acknowledgments}

\appendix

\section{Neural Ordinary Differential Equation} \label{SecI}
The NODE \cite{Chen2019} inherits the basic idea of the residual network \cite{He2016} and is able to reconstruct the differential equations satisfied by certain sequential data $\mathcal{D}$. The NODE makes the ansatz $\dot{\boldsymbol{\rho}} = \boldsymbol{f}[\boldsymbol{\rho}(t),\boldsymbol{\theta},t]$.
If the specific form of $\boldsymbol{f}$ is priorly unknown, a deep neural network can be used to establish the mapping from $\{\boldsymbol{\rho}(t),\boldsymbol{\theta},t\}$ to $\boldsymbol{f}$. However, if certain prior information is available, e.g., $\boldsymbol{f}$ satisfies the Lindblad equation as the situation considered in this work, the question comes to determine the unknown parameters $\boldsymbol{\theta}$ in $\boldsymbol{f}$. Given an initial state $\boldsymbol{\rho}(t_0)$, one can propagate the NODE forwardly from $t_0$ to $t_N$ using certain ODE solvers and obtain predictive solutions $\{\dot{\boldsymbol{\rho}}_\text{pred}(t_n)\}$. The loss function $\mathcal{L}$ is defined as the effective distance between the $\{\dot{\boldsymbol{\rho}}_\text{pred}(t_n)\}$ and the realistic values $\{\dot{\boldsymbol{\rho}}_\text{real}(t_n)\}\in\mathcal{D}$. Next, we show how to obtain the derivative $d\mathcal{L}/d\boldsymbol{\theta}$, which is based on the the adjoint field method \cite{Chen2019}. 

The augmented adjoint state $\boldsymbol{a}_\text{aug}$ is defined as the derivative of $\mathcal{L}$ with respect to the augmented states $\boldsymbol{\rho}_\text{aug} = \left[ \boldsymbol{\rho}, \boldsymbol{\theta}, t \right]^T$, i.e.,
\begin{equation}
\begin{aligned}
\boldsymbol{a}_\text{aug} &= \frac{\partial \mathcal{L}}{\partial \boldsymbol{\rho}_\text{aug}} \\
&= \left[\frac{\partial \mathcal{L}}{\partial\boldsymbol{\rho}(t)}, \frac{\partial \mathcal{L}}{\partial\boldsymbol{\theta}(t)}, \frac{\partial \mathcal{L}}{\partial t}\right]^T \\
&= \left[\boldsymbol{a},\boldsymbol{a}_\theta,a_t\right]^T,
\end{aligned}
\label{S2}
\end{equation}
which satisfies the differential equation
\begin{equation}
\begin{aligned}
\frac{d \boldsymbol{a}_\text{aug}^T}{dt} &= -\boldsymbol{a}_\text{aug}^T \cdot \frac{\partial \boldsymbol{f}_\text{aug}}{\partial \boldsymbol{\rho}_\text{aug}^T},
\end{aligned}
\label{S4}
\end{equation}
where
\begin{equation}
\boldsymbol{f}_\text{aug} = \frac{d \boldsymbol{\rho}_\text{aug}}{d t}
= \left[ \dot{\boldsymbol{\rho}}(t), \dot{\boldsymbol{\theta}}, \dot{t} \right]^T
= \left[ \boldsymbol{f}, \mathbf{0}, 1 \right]^T
\label{S5}
\end{equation}
is the derivative of the augmented state with respect to $t$, and
\begin{equation}
\frac{\partial \boldsymbol{f}_\text{aug}}{\partial \boldsymbol{\rho}_\text{aug}^T} = \frac{\partial \left[ \boldsymbol{f}, \mathbf{0}, 1 \right]}{\partial \left[ \boldsymbol{\rho}, \boldsymbol{\theta}, t \right]^T}=
\begin{bmatrix}
\frac{\partial \boldsymbol{f}}{\partial \boldsymbol{\rho}} & \frac{\partial \boldsymbol{f}}{\partial \boldsymbol{\theta}} & \frac{\partial \boldsymbol{f}}{\partial t}\\
\mathbf{0} & \mathbf{0} & \mathbf{0}\\
\mathbf{0} & \mathbf{0} & 0\\
\end{bmatrix}
\label{S6}
\end{equation}
is the Jacobean matrix. 
Substituting Eqs.~(\ref{S6}) and (\ref{S2}) into Eq.~(\ref{S4}) we have
\begin{equation}
\left[\dot{\boldsymbol{a}}^T,\dot{\boldsymbol{a}}_\theta^T,\dot{a_t}\right]= -\left[\boldsymbol{a}^T \cdot \frac{\partial \boldsymbol{f}}{\partial \boldsymbol{\rho}^T}, \boldsymbol{a}^T \cdot \frac{\partial \boldsymbol{f}}{\partial \boldsymbol{\theta}^T},\boldsymbol{a}^T \cdot \frac{\partial \boldsymbol{f}}{\partial t}\right],
\label{S8}
\end{equation}
and through backwardly propagating which from $t_N$ to $t_0$ we obtain the derivative $\boldsymbol{a}_\theta(t_0)=d\mathcal{L}/d \boldsymbol{\theta}|_{t=t_0}$. Particularly, the boundary condition for the backward propagation is given by
\begin{eqnarray}
\boldsymbol{a}(t_N) &=& \frac{\partial\mathcal{L}}{\partial \boldsymbol{\rho}(t_N)}, \notag\\
\boldsymbol{a}_\theta(t_N)&=& \left.\frac{\partial\mathcal{L}}{\partial \boldsymbol{\theta}}\right|_{t=t_N} = \left.\left(\frac{\partial\mathcal{L}}{\partial \boldsymbol{\rho}^T}\cdot\frac{\partial \boldsymbol{\rho}}{\partial t}\right)\frac{\partial t}{\partial \boldsymbol{\theta}}\right|_{t=t_N} = \mathbf{0}, \\
a_t(t_N)&=& \left.\frac{\partial\mathcal{L}}{\partial t}\right|_{t=t_N} = \left.\frac{\partial\mathcal{L}}{\partial \boldsymbol{\rho}^T}\cdot\frac{\partial \boldsymbol{\rho}}{\partial t}\right|_{t=t_N} = \boldsymbol{a}^T(t_N) \cdot \boldsymbol{f}(t_N),  \notag
\label{S9}
\end{eqnarray}
where $\partial\mathcal{L}/\partial \boldsymbol{\rho}(t_N)$ can be directly calculated using the automatic differentiation toolbox \cite{Baydin2018, Paszke2017}.

Note that, for a sequential data with intermediate data points at $t_n$, the total loss function $\mathcal{L}$ is a summation of individual $\mathcal{L}_n$, and hence $d\mathcal{L}/d\boldsymbol{\theta}$ is simply the average over all the periods of back propagation from $t_n$ to $t_{n-1}$, i.e., $d\mathcal{L}/d\boldsymbol{\theta} = N_\text{ts}^{-1} \sum_{n=0}^{N_\text{ts}-1} \boldsymbol{a}_\theta(t_n)$.

\section{Complementary Result of the Two Examples} \label{SecII}
Here, we provide more information on the two examples.
The 4-by-4 Hamiltonian $H$ of the spin-3/2 model can be expanded by the SU(4) Hermitian generators $\hat{g}_\mu$ in the fundamental representation. We practically take the matrices as \cite{Bertlmann2008}
\begin{widetext}
\begin{equation}
\begin{aligned}
\hat{g}_1 &=
\begin{pmatrix}
0 & 1 & 0 & 0 \\
1 & 0 & 0 & 0 \\
0 & 0 & 0 & 0 \\
0 & 0 & 0 & 0
\end{pmatrix}, \
\hat{g}_2 =
\begin{pmatrix}
0 & -i & 0 & 0 \\
i & 0 & 0 & 0 \\
0 & 0 & 0 & 0 \\
0 & 0 & 0 & 0
\end{pmatrix}, \
\hat{g}_3 =
\begin{pmatrix}
1 & 0 & 0 & 0 \\
0 & -1 & 0 & 0 \\
0 & 0 & 0 & 0 \\
0 & 0 & 0 & 0
\end{pmatrix},  \
\hat{g}_4 =
\begin{pmatrix}
0 & 0 & 1 & 0 \\
0 & 0 & 0 & 0 \\
1 & 0 & 0 & 0 \\
0 & 0 & 0 & 0
\end{pmatrix},  \
\hat{g}_5 =
\begin{pmatrix}
0 & 0 & -i & 0 \\
0 & 0 & 0 & 0 \\
i & 0 & 0 & 0 \\
0 & 0 & 0 & 0
\end{pmatrix},
\\
\hat{g}_6 &=
\begin{pmatrix}
0 & 0 & 0 & 0 \\
0 & 0 & 1 & 0 \\
0 & 1 & 0 & 0 \\
0 & 0 & 0 & 0
\end{pmatrix},
\hat{g}_7 =
\begin{pmatrix}
0 & 0 & 0 & 0 \\
0 & 0 &-i & 0 \\
0 & i & 0 & 0 \\
0 & 0 & 0 & 0
\end{pmatrix},
\hat{g}_8 =
\frac{1}{\sqrt{3}}\begin{pmatrix}
1 & 0 & 0 & 0 \\
0 & 1 & 0 & 0 \\
0 & 0 & -2 & 0 \\
0 & 0 & 0 & 0
\end{pmatrix},
\hat{g}_9 =
\begin{pmatrix}
0 & 0 & 0 & 1 \\
0 & 0 & 0 & 0 \\
0 & 0 & 0 & 0 \\
1 & 0 & 0 & 0
\end{pmatrix},
\hat{g}_{10} =
\begin{pmatrix}
0 & 0 & 0 & -i \\
0 & 0 & 0 & 0 \\
0 & 0 & 0 & 0 \\
i & 0 & 0 & 0
\end{pmatrix},
\\
\hat{g}_{11} &=
\begin{pmatrix}
0 & 0 & 0 & 0 \\
0 & 0 & 0 & 1 \\
0 & 0 & 0 & 0 \\
0 & 1 & 0 & 0
\end{pmatrix},
\hat{g}_{12} =
\begin{pmatrix}
0 & 0 & 0 & 0 \\
0 & 0 & 0 & -i \\
0 & 0 & 0 & 0 \\
0 & i & 0 & 0
\end{pmatrix},
\hat{g}_{13} =
\begin{pmatrix}
0 & 0 & 0 & 0 \\
0 & 0 & 0 & 0 \\
0 & 0 & 0 & 1 \\
0 & 0 & 1 & 0
\end{pmatrix},
\hat{g}_{14} =
\begin{pmatrix}
0 & 0 & 0 & 0 \\
0 & 0 & 0 & 0 \\
0 & 0 & 0 & -i \\
0 & 0 & i & 0
\end{pmatrix},
\hat{g}_{15} =
\frac{1}{\sqrt{6}}\begin{pmatrix}
1 & 0 & 0 & 0 \\
0 & 1 & 0 & 0 \\
0 & 0 & 1 & 0 \\
0 & 0 & 0 & -3
\end{pmatrix}.  \\
\end{aligned}
\end{equation}
\end{widetext}
\begin{figure*}[t]
\includegraphics[width=0.96\textwidth]{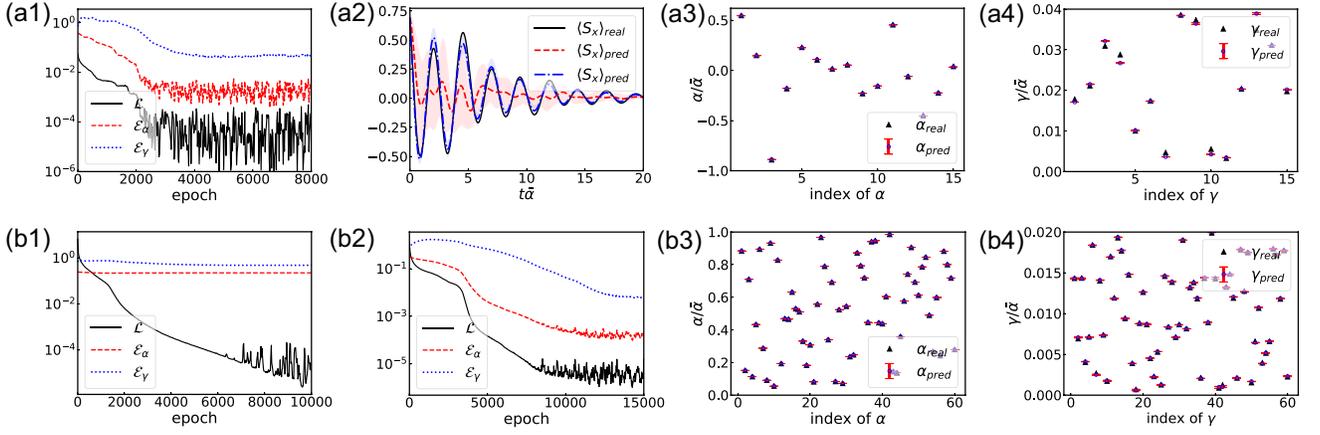}
\caption{
First row: learning results for the spin-3/2 model with the t-independent probe. (a1) Dependence of $\mathcal{L}$, $\alpha$ and $\gamma$ on training epochs. (a2) Spin dynamics of $\langle S_x \rangle_\text{real} \in \mathcal{D}$, $\langle S_x \rangle_\text{pred}$ before training and $\langle S_x \rangle_\text{pred}$ after $10^3$ epochs of training. (a3) and (a4) Real parameters (solid triangles) and predictive parameters (circles with error bars) after training. Second row: learning results for the spin-1/2 chain with the t-dependent probe. (b1) Training processes of $\mathcal{L}$, $\alpha$ and $\gamma$ with the dataset being constructed by total spin measurements, i.e. $\mathcal{D}=\{\langle S_{\mu=x,y,z} (\mathcal{P}_s,t_n)\rangle \}$. (b2)-(b4) Learning results on the dataset constructed by local spin measurements, i.e. $\mathcal{D}=\{\langle \sigma_{\mu=x,y,z}^{i=3} (\mathcal{P}_s,t_n) \rangle\}$.
}
\label{FigS2}
\end{figure*}
These generators naturally satisfy the traceless condition $\text{Tr}[\hat{g}_\mu]=0$ and the orthogonal condition $\text{Tr}[\hat{g}_\mu \hat{g}_\nu]=2\delta_{\mu\nu}$, which are also the properties that are supposed to be satisfied by the Lindblad dissipative operators $J_k$. One may check that the one-body and the two-body dissipative operators of the second example (spin-1/2 chain) also satisfy these two conditions. Both conditions would be broken if any $J_k$ is with a finite trace $a_k = \text{Tr}[\hat{g}_\mu]\neq0$. However, since the Lindblad equation Eq.~(\ref{GKSLE}) is invariant under the transformation
\begin{equation}
\begin{aligned}
J_k &\longrightarrow J_k - a_k, \\
H &\longrightarrow H + \frac{i}{2}\sum_\mu[a_k^* J_k - (a_k^* J_k)^\dagger],
\end{aligned}
\label{S12}
\end{equation}
one can always recover the conditions by absorbing $a_k$ into $H$.
Note that, the Lindblad equation has another invariance $J_k \rightarrow e^{i \phi_k} J_k$ with $\phi_k$ being an arbitrary phase factor, which implies that the phase of $\gamma_k$ is meaningless, and hence we set $\gamma_k$ positive throughout this work.

The first row of Fig.~\ref{FigS2} shows the learning result of the spin-3/2 model using the t-independent probe, where subfigures denote the dependence of $\mathcal{L}$, $\mathcal{E}_\alpha$ and $\mathcal{E}_\gamma$ on training epochs (a1), the spin dynamics before and during training (a2), and the predictive parameters $\alpha$ (a3) and $\gamma$ (a4) after training, respectively. It is indicated that the NODE can also accurately reproduce the spin-3/2 model with predictive errors $\mathcal{E}_\alpha \lesssim 1\text{\textperthousand}$ and $\mathcal{E}_\gamma \lesssim 5\%$. The second row of Fig.~\ref{FigS2} displays the result with respect to the spin-chain model using the t-dependent probe. Specifically in Fig.~\ref{FigS2}(b1), the data points were collected on the measurement of total spins. There, the loss function $\mathcal{L}$ decreases but the predictive errors $\mathcal{E}_{\alpha,\gamma}$ barely decrease. In contrast, if the measurement is performed on the local spins, favorable results are obtained as is illustrated in the Figs.~\ref{FigS2}(b2)-(b4). These behaviors verify the "symmetry" remarks discussed in the main text, since the local observations $\langle \sigma_\mu^i \rangle$ explicitly break the translational "symmetry".

\section{General Solution of Lindblad Equation}  \label{SecIII}
We obtain the solution of the Lindblad equation by mapping the Liouvilian operator $L$ into the double Hilbert space, which is generally known as the Choi-Jamiolkwski isomorphism \cite{Choi1975,Jamiolkowski1972} or vectorization, i.e.,
\begin{equation}
\begin{aligned}
\tilde{L} =& -i[\mathbb{I} \otimes H-H^T\otimes \mathbb{I} ] \\
&+ \sum_k\left[J_k^*\otimes J_k - \frac{1}{2}\left(\mathbb{I} \otimes J_k^\dagger J_k + (J_k^\dagger J_k)^T \otimes \mathbb{I} \right)\right],
\end{aligned}
\label{S13}
\end{equation}
where $\tilde{L}$ is the vectorized Liouvilian operator and $\mathbb{I}$ is an identity operator with the same shape as that of $H$. Generally, $\tilde{L}$ is non-Hermitian with the spectrum structure $\tilde{L} = \sum_m \epsilon_m \mathbf{u}_m \cdot \mathbf{v}^\dagger_m$ where $\epsilon_m$ denotes the Liouvilian spectrum, and $\mathbf{u}_m$ and $\mathbf{v}_m$ are the right and the left eigenvectors, respectively.
In such a framework, the general solution of the Lindblad equation can be obtained as
\begin{equation}
\tilde{\boldsymbol{\rho}}(t) = \sum_m \lambda_m e^{\epsilon_m t} \mathbf{u}_m,
\label{S16}
\end{equation}
where $\tilde{\boldsymbol{\rho}}$ is the vectorized density operator, and
\begin{equation}
\lambda_m = \mathbf{v}^\dagger_m \cdot \tilde{\boldsymbol{\rho}}(t_0)
\label{S17}
\end{equation}
characterizes the projective coefficients with respect to the initial state $\tilde{\boldsymbol{\rho}}(t_0)$.

The Liouvilian spectrum $\epsilon_m = \epsilon_m^r + i \epsilon_m^i$ is generally complex, where the imaginary part $\epsilon_m^i$ characterizes the undamped oscillations, while the real part $\epsilon_m^r\leq0$ accounts for the damping of $\mathbf{u}_m$ during evolution. In the spectrum, it is well-known that there exists an undamped state $\mathbf{u}_0$ with $\epsilon_0 =0$, which is the steady state to which the system relaxes after a long evolution. The energy gap $\Delta = \epsilon_0 - \epsilon_1 = \Delta_1 + i \Delta_2$ between the slowest damping state $\mathbf{u}_1$ and the steady state $\mathbf{u}_0$ is the Liouvilian gap. The real gap $\Delta_1>0$ determines the decoherent time in the way of $t_\text{dc} = 1/\Delta_1$. For weak dissipation $\bar{\gamma} \ll \bar{\alpha}$, $\Delta_1$ is linearly proportional to the dissipative strength $\bar{\gamma}$, i.e., $\Delta_1=-C\bar{\gamma}$, with $C$ a non-universal factor depending on the number of dissipation channels and the particular form of external probes, whereas the imaginary gap $\Delta_2$ is characterized by the energy scale of $\bar{\alpha}$. Hence, we generally have $\Delta_1 \ll |\Delta_2|$.

The dataset $\mathcal{D}$ is constructed by the measurement
\begin{equation}
\begin{aligned}
\left \langle O(t) \right \rangle &=\sum_m \lambda_m e^{\epsilon_m t} \tilde{O}^\dagger \cdot \mathbf{u}_m \\
&\approx \lambda_0 (\tilde{O}^\dagger \cdot \mathbf{u}_0) + \lambda_1 e^{-\Delta_1 t} f(t), \\
\end{aligned}
\label{S18}
\end{equation}
where $\tilde{O}$ is the vectorized observable $O$ and $f(t) = e^{- i \Delta_2 t} (\tilde{O}^\dagger \cdot \mathbf{u}_1) + \text{c.c.}$ In the second line, we have neglected the contributions of faster damping states with $m>1$. Clearly, the first term is a t-independent constant; the second term exhibits a fast oscillation $f(t)$ modulated by a slowly damped envelop $e^{-\Delta_1 t}$.
Taking derivative of Eq.~(\ref{S18}) with respect to $\bar{\gamma}$ leads to $\chi(t)$ [Eq.~(\ref{Sus})], based on which one can obtain $\eta$ as $\bar{\gamma}$ approaches the realistic value $\bar{\gamma}_\text{real}$. Particularly for the MAE loss $\eta_\text{MAE}$, we have
\begin{equation}
\begin{aligned}
\eta_\text{MAE} &= \frac{1}{N}\sum_n \left|\frac{d |\langle O(t_n) \rangle_\text{pred} - \langle O(t_n) \rangle_\text{real}|}{d \bar{\gamma}}\right| \\
&= \frac{1}{N}\sum_n \left|\frac{d \langle O(t_n) \rangle_\text{pred}}{d \bar{\gamma}}\right| \\
&= \frac{1}{N}\sum_n \chi(t).
\end{aligned}
\label{S21}
\end{equation}
On the other hand, for the MSE loss function, $\eta_\text{MSE}$ can be calculated by
\begin{equation}
\begin{aligned}
\eta_\text{MSE} &= \frac{1}{N}\sum_n \left|\frac{d \left(\langle O(t_n) \rangle_\text{pred} - \langle O(t_n) \rangle_\text{real}\right)^2}{d \bar{\gamma}}\right| \\
&= \frac{1}{N}\sum_n \left| \langle O(t_n) \rangle_\text{pred} - \langle O(t_n) \rangle_\text{real} \right| \left|\frac{d \langle O(t_n) \rangle_\text{pred}}{d \bar{\gamma}}\right| \\
&= \frac{1}{N}\sum_n |\delta \bar{\gamma}|\left| \frac{\langle O(t_n) \rangle_\text{pred} - \langle O(t_n) \rangle_\text{real}}{\delta \bar{\gamma}} \right| \left|\frac{d \langle O(t_n) \rangle_\text{pred}}{d \bar{\gamma}}\right| \\
&\propto \frac{1}{N}\sum_n \chi(t)^2,
\end{aligned}
\label{S23}
\end{equation}
with $\delta \bar{\gamma} = \bar{\gamma} - \bar{\gamma}_\text{real}$. Eqs. (\ref{S21}) and (\ref{S23}) reproduce the Eq.~(\ref{EPs}).


\begin{thebibliography}{99}
\bibitem{Complexity} To deal with an open quantum system with $L$ spins is equivalent to deal with a closed system with $2L$ spins, since the open system can be viewed as two coupled closed systems under the Choi-Jamiolkwski isomorphism [see Eq.~(\ref{S13}) in Appendix \ref{SecIII}]. The computational complexity are further enlarged by $N_\text{bs} N_\text{ini}\sim 10^2$ times due to batches and initializations. Practically, we rent two Nvidia GTX 1080 Ti graphics cards to speed up the numerics, which keeps the training time of the spin-chain task with $L=5$ at about one day.
\end{thebibliography}
\end{document}